\documentclass[a4paper, 11pt]{article}

\usepackage[utf8x]{inputenc}
\usepackage{epsfig,amssymb,euscript,xspace, color}
\usepackage{jheppub}
\usepackage[T1]{fontenc}
\usepackage{color}
\usepackage{amssymb}
\usepackage{amsmath}
\usepackage{graphicx}
\usepackage{mathtools}
\usepackage{ragged2e}

\usepackage{dcolumn}
\usepackage{graphicx}
\usepackage{braket}
\usepackage{verbatim} 
\usepackage[english]{babel}
\usepackage{hyperref}
\hypersetup{
citecolor=red,
colorlinks=true,
filecolor=red,
linkcolor=blue,
linktocpage=true,
urlcolor=blue
} 
\usepackage[titletoc,toc,title]{appendix}
\numberwithin{equation}{section}
\usepackage{cleveref}
\usepackage{epsfig}
\usepackage{float}
\usepackage{amsfonts}
\usepackage{enumitem}
\usepackage[font={footnotesize,it}]{caption}

\newcommand{\be}{\begin{equation}}
\newcommand{\ee}{\end{equation}}
\newcommand{\eq}[1]{(\ref{#1})}

\def\b{\beta}

\def\d{\delta}
\def\D{\Delta}
\def\e{\epsilon}

\def\r{\rho}

\def\t{\tau}

\def\({\left(}
\def\){\right)}
\def\[{\left[}
\def\]{\right]}

\def\wb{{\bar w}}

\def \be {\begin{equation}}
\def \ee {\end{equation}}
\def \ba {\begin{array}}
\def \ea {\end{array}}
\def \bea{\begin{eqnarray}}
\def \eea{\end{eqnarray}}

\def \b {\beta}

\def \d {\delta}
\def \D {\Delta}
\def \e {\epsilon}

\def \r {\rho}

\def \t {\tau}

\def\bea{\begin{eqnarray}}
\def\eea{\end{eqnarray}}
\newcommand{\bit}{\begin{itemize}}  \newcommand{\eit}{\end{itemize}}
\newcommand{\ben}{\begin{enumerate}}  \newcommand{\een}{\end{enumerate}}

\def\cA{{\cal A}}  
  
\def\cG{{\cal G}}

\title{Time-like Entanglement Entropy in AdS/BCFT}

\author[a,b,c]{Chong-Sun Chu}
\author[b,c]{Himanshu Parihar}

\affiliation[a]{
	Department of Physics, National Tsing-Hua University,
 Hsinchu 30013, Taiwan}
\affiliation[b]{Center of Theory and Computation,
National Tsing-Hua University,
 Hsinchu 30013, Taiwan}
\affiliation[c]{Physics Division,
    National Center for Theoretical Sciences,
     Taipei 10617, Taiwan}

\emailAdd{cschu@phys.nthu.edu.tw}
\emailAdd{himansp@phys.ncts.ntu.edu.tw}

\abstract{\noindent We study the entanglement entropy for time-like subsystem in
two-dimensional
boundary conformal field theory (BCFT) both from the
field theory and holographic point of view. In field theory, we compute the
time-like entanglement entropy of a pure time-like interval at zero
and finite temperature using the replica technique and
analytical continuation.
We find that, similar to the ordinary space-like entanglement entropy
in BCFT, the time-like entropy also has a bulk phase and a boundary
phase which corresponds respectively
to the dominance of the identity block
in the bulk and boundary OPE channels. However, we find that in
Lorentzian BCFT, the
time-like entanglement entropy  posses a third {\it Regge
  phase} which arises in the Regge limit of the interval,
when one endpoint of the time interval
approaches the light cone of the mirror image of the other endpoint.
We determine the phase diagram for the time-like entanglement entropy.
We find that while the time-like entropy
is complex in the bulk phase and has a boundary
term in the boundary phase, there is no boundary
entropy in the Regge phase. Moreover,
it can be real or complex depending on which
side the Regge limit is approached from.
On the gravity side, we obtain the holographic
time-like entanglement entropy from the corresponding bulk dual
geometries  and find exact agreement with the field theory results.
The time-like entanglement entropy may be useful to describe
the entanglement of a quantum dot on a half line. }

\begin{document} 
	
	\maketitle
	\flushbottom
	\pagebreak

	\definecolor{orange}{rgb}{1.0, 0.49, 0.0}

\section{Introduction}\label{Intro}

\justify

The AdS/CFT correspondence \cite{Maldacena:1997re} provides an
interesting relation between the $d$ dimensional conformal field
theory (CFT) to that of quantum gravity on a $(d+1)$ asymptotically
anti-de Sitter (AdS) spacetime. One can extend the AdS/CFT to the case
where a CFT is defined on a manifold with boundaries (e.g. CFT on a
half plane). This class of CFTs with a boundary is called a boundary
conformal field theory (BCFT) when a part of conformal symmetry
(called boundary conformal symmetry) is preserved at the boundary
\cite{Cardy:1984bb,Cardy:1989ir,McAvity:1993ue,McAvity:1995zd,Cardy:2004hm}.
The holographic dual of such BCFTs are given by an asymptotically AdS
spacetime truncated by an end of the world (EOW) brane which satisfy
either the Neumann boundary condition \cite{Takayanagi:2011zk,Fujita:2011fp}
as it was originally proposed, the conformal boundary condition
\cite{Miao:2017gyt,Chu:2017aab} or the Dirichlet boundary condition \cite{Miao:2018qkc} which were
found later to all define consistent holographic BCFT.

Owed to discovery of the  Ryu-Takayanagi (RT) formula \cite{Ryu:2006bv,Ryu:2006ef} for the entanglement entropy for  bipartite pure states, 
the holographic study of  quantum entanglement has become one of the most
active recent developments of AdS/CFT. 
The entanglement entropy for such bipartite states can be
obtained in a $(1+1)$ dimensional conformal field theories using a
replica technique
\cite{Calabrese:2004eu,Calabrese:2009qy}.
The holographic entanglement entropy of a
subregion $A$ in a dual CFT may be computed in terms of the area of
the codimension two extremal surface ($\Gamma_A$) homologous to the
boundary subregion as 
\begin{equation}
S_A=\frac{\mathrm{Area (\Gamma_A)}}{4G_N}.
\end{equation}
This formula becomes more interesting in AdS/BCFT since the
minimal surface can now end on the EOW brane as first noted in
\cite{Takayanagi:2011zk,Fujita:2011fp}. Moreover, depending on the
position of the interval $A$  from the boundary, there will be competition
between the bulk minimal surface and the minimal surface that ends on the
EOW brane and give rises to a phase transition
of the entanglement entropy since the two surfaces have different topology
\cite{Miao:2017gyt,Chu:2017aab}.
In field theory, this corresponds to
the shift from the vacuum block dominance in the OPE of the bulk channel OPE 
to the boundary channel \cite{Sully:2020pza}.

Recently, the authors of \cite{Doi:2022iyj} proposed a new quantity
for the entanglement measure for a time-like subsystem known as the
\textit{time-like entanglement entropy}\footnote{Note that it has been introduced earlier in the context of $T\bar{T}$ deformed version of the AdS$_3$/CFT$_2$ correspondence in \cite{Wang:2018jva}.}. It is defined by analytically
continuing the entanglement entropy of a space-like subsystem $A$ to
the case where the subsystem $A$ becomes a time-like subsystem. The
time-like entanglement entropy (TEE) is found to have a complex
value in a CFT$_2$. Furthermore, it was shown that the extremal
surface for the holographic TEE consists of space-like and
time-like parts which give rises to the real part and the imaginary
part of the time-like entanglement entropy respectively.
It was also argued that \cite{Doi:2022iyj,Doi:2023zaf} the TEE is a
special example of pseudo entropy
\cite{Nakata:2020luh,Mollabashi:2020yie,Mollabashi:2021xsd}.  See
\cite{Liu:2022ugc,Diaz:2021snw,Guo:2022jzs,Aalsma:2022swk,Narayan:2022afv,
  Reddy:2022zgu,Li:2022tsv,Parzygnat:2022pax,He:2023eap,Alshal:2023kcd,
  Cotler:2023xku,Chen:2023prz,Chen:2023gnh,Jiang:2023ffu,Narayan:2023ebn,Jiang:2023loq}
for recent progress related to TEE.

The above developments lead naturally to an interesting question about the
time-like entanglement entropy in BCFT. In fact,
from the point of view of BCFT and replica technique,
the bulk phase and boundary phase of the entanglement entropy arises from
the vacuum block dominance in the bulk and boundary channel of the OPE of the
twist operators, which occurs in the bulk limit ($\xi \to 0$)
and the boundary limit $(\xi \to \infty$) respectively.
Here $\xi$ is the cross ratio
for the two endpoints of the interval together with their images.
Now in a Lorentzian BCFT, the two point function
is known to have a branch point singularity  in the Regge limit $\xi \to -1$
\cite{Mazac:2018biw}, which arises when one endpoint of the
time interval approaches the light-cone of the
mirror image of the other endpoint\footnote{
We remark that the Regge limit, and hence the Regge phase of the TEE,
is possible only in BCFT when the interval is time-like.}.
Therefore, in addition to the bulk and boundary limits, one can expect to
find new interesting properties for the TEE in the Regge limit of BCFT.
The study of this richer phase structure of TEE in BCFT
is one of the main motivations of this paper. We find that
unlike the usual
space-like entropy which consists of a bulk phase and a boundary phase, the
TEE in BCFT has a new third phase 
which arises in the Regge limit of the time-like interval. An interesting
novel feature here is that the TEE goes from real value to complex
as the time interval goes pass the Regge limit point.

This article is organized as follows. In \cref{TEE-cft-review} we
briefly review about the time-like entanglement entropy in
AdS${}_3$/CFT${}_2$. In \cref{{review-bcft}} we review some of
the salient features of BCFT. In
\cref{TEE-bcft} we obtain the time-like entanglement entropy of a pure time-like
interval at zero and finite temperature in a
BCFT${}_2$. We identify the Regge phase and determine the phase diagram
for the TEE.
In \cref{HTEE} we compute the 
time-like entanglement entropy using AdS${}_3$/BCFT${}_2$.  
The RT
surface in the bulk phase consists of space-like and time-like
geodesics leading to a complex valued TEE. In the boundary phase, the RT
surface consists of two geodesics ending on the EOW brane and it gives
a real valued TEE. Finally, the RT surface in the Regge phase is
given by the geodesic joining the end point of interval and ending on
the plane perpendicular to the boundary.
The geodesics lie along the plane crossing one end point of the
interval and mirror reflection of the other end point of the
interval. Remarkably, we obtain exact matches between the field theory
replica technique results and the bulk holographic computation of the
TEE. Finally in \cref{Summary} we present our summary
and discussions.

\section{Time-like entanglement entropy in AdS${}_3$/CFT${}_2$}
\label{TEE-cft-review}
In this section, we briefly review the time-like entanglement entropy
for the configuration of a time-like interval in AdS${}_3$/CFT${}_2$
framework \cite{Doi:2022iyj}. Consider a generic space-like interval
$A\equiv [(t_1,x_1),(t_2,x_2)]$ in a CFT$_2$ whose time-like and
space-like width are given by $t_{12}=T_0$ and $x_{12}=X_0$
respectively. Then, the entanglement entropy $S_A$ of an interval $A$
may be obtained using the replica technique as
\cite{Calabrese:2004eu,Calabrese:2009qy}
\begin{equation}\label{ent-gen}
S_A=\frac{c}{3}\log\frac{\sqrt{X_0^2-T_0^2}}{\epsilon},
\end{equation}
where $\epsilon$ is a UV cut-off and $c$ is the central charge of the CFT.  The time-like entanglement entropy $S_A^T$ for a purely
time-like interval $A$ is obtained by analytically continuing the
space-like interval to a time-like interval followed by taking $X_0=0$
in \cref{ent-gen} as follows \cite{Doi:2022iyj}
\begin{equation}\label{TEE-cft}
S_A^T=\frac{c}{3}\log\left(\frac{T_0}{\epsilon}\right)+\frac{i\pi c}{6}.
\end{equation}
It is observed that the above TEE takes complex values for the
standard unitary CFTs.

At finite temperature $T=1/\beta$, 
the entanglement entropy for a generic space-like
interval of width $t_{12}$ and $x_{12}$ at finite temperature is
obtained by evaluating the two point twist correlator on a cylinder of
circumference $\beta$
and the result is
\cite{Calabrese:2004eu,Calabrese:2009qy}
\begin{align}\label{EE-T}
  S_A=\frac{c}{6} \log\left[\frac{\beta^2}{\pi^2\epsilon^2}
    \sinh\left(\frac{\pi}{\beta}(x_{12}+ t_{12})\right)\sinh\left(\frac{\pi}
              {\beta}(x_{12}-t_{12})\right)\right].
\end{align}
Then, the time-like entanglement entropy for a purely time-like interval at
finite temperature can be computed by taking $x_{12}=0$ and $t_{12}=T_0$
in \cref{EE-T} as follows \cite{Doi:2022iyj}

\begin{align}\label{TEE-T-cft}
  S_A^T=\frac{c}{3}\log\left[\frac{\beta}{\pi\epsilon}\sinh\frac{\pi T_0}
    {\beta}\right]+\frac{i\pi c}{6}.
\end{align}
The above result for  TEE at finite temperature also have complex value
similar to the zero temperature case.

Next we review the holographic time-like entanglement entropy of a
time-like interval in the AdS${}_3$/CFT${}_2$ scenario
\cite{Doi:2022iyj}. To this end, consider the Poincar\'{e} patch of a
AdS$_3$ spacetime (with AdS radius $R_\mathrm{AdS}=1$) whose metric is given
by

\begin{equation}
ds^2=\frac{dz^2-dt^2+dx^2}{z^2}.
\end{equation}
The time-like interval $A$ of the boundary CFT$_2$ has finite width
$T_0$ along the time direction and fixed spatial coordinate in $x$
direction described by $x_1=x_2$. It is argued in \cite{Doi:2022iyj}
that due to the lack of space-like geodesic connecting the boundary of
time-like interval $A$, one should use two space-like geodesics
connecting the endpoints of $A$ and null infinities followed by a
time-like geodesic which connects the endpoints of two space-like
geodesics. The holographic time-like entanglement entropy for the
time-like interval is then given by the length of these three
geodesics.  The equation for the space-like geodesic which connects
endpoints $\partial A$ with null infinities is given by
\begin{equation}
t=\sqrt{z^2+T_0^2/4},
\end{equation}
which can identified as a semi-circle geodesic via the Wick
rotation. On utilizing the RT formula, the time-like
entanglement entropy is given by the length of this space-like
geodesic as
\begin{equation}
  S_A^T=\frac{1}{4G_N} 2T_0\int^{\infty}_{\epsilon}
  \frac{dz}{2z\sqrt{z^2+T_0^2/4}}=\frac{c}{3}\log\frac{T_0}{\epsilon},
\end{equation}
where $c=\frac{3}{2 G_N}$ 
is used \cite{Brown:1986nw}.
It matches with the real part of the dual CFT$_2$ result in \cref{TEE-cft}.
The imaginary part of the time-like entanglement entropy is obtained by
embedding the Poincar\'{e} patch in the global patch whose metric is given by
\begin{equation}
ds^2=-\cosh^2\rho d\tau^2+d\rho^2+\sinh^2\rho d\theta^2.
\end{equation}
Then, the imaginary part is given by the length of a time-like
geodesic connecting two endpoints at $\rho=0$ and
$\tau=\pm\frac{\pi}{2}$ in the global coordinates. The length of this
time-like geodesic is $\pi c/6$ which matches with the imaginary part
of dual CFT$_2$ result as described in \cref{TEE-cft}.

For the time-like interval $A$ of length $T_0$ at a finite
temperature, the gravity dual is described by the planar BTZ black
hole whose metric (with $R_{AdS}=1$) is given by
\begin{equation}\label{btz-metric}
ds^2=-\left(r^2-r_{+}^2\right)dt^2+\frac{dr^2}{r^2-r_{+}^2}+r^2dx^2,
\end{equation}
where $r_+=\frac{2\pi}{\beta}$ and $\beta$ is the inverse temperature
of the dual CFT$_2$.  Similarly, the time-like entanglement entropy
for a time-like interval $A$ at a finite temperature is given by the
length of space-like and time-like extremal surfaces in the BTZ black
hole background which can be expressed as \cite{Doi:2022iyj}
\begin{equation}\label{tee-finte-temp-cft}
  S_A^T=\frac{c}{3}\log\left[\frac{\beta}{\pi\epsilon}
    \sinh\left(\frac{\pi T_0}{\beta}\right)\right]+\frac{i \pi c}{6},
\end{equation}
which agrees with the corresponding dual field theory result (\ref{TEE-T-cft}).

\section{Review of BCFT}\label{review-bcft}
In this section, we will briefly review the salient features of the
boundary conformal field theories
\cite{Cardy:1984bb,Cardy:2004hm,Mazac:2018biw,Reeves:2021sab}. A
boundary conformal field theory (BCFT) is a conformal field theory
(CFT) defined on a manifold with boundaries such that a part of
conformal symmetry is preserved by the boundaries. There are many
different possible choices of conformal-invariant boundary
conditions for a given CFT.  When the BCFT$_d$ lives on the half-plane
$\mathbb{R}^{d-1} \times \mathbb{R}_+$ with a planar boundary, a
BCFT$_d$ has the symmetry $SO(d- 1,2)$ in Lorentzian signature (or
$SO(d,1)$ in Euclidean signature) which is the subgroup of original
CFT$_d$ symmetry group.

Consider a
BCFT in $d$ dimension space with a time independent planar boundary.
Denote the coordinates by  
$x=(x_0, \vec{x}, x_\perp )$ where
$x_0$ is the time coordinate, 
$\vec{x}$ are the Euclidean coordinates parallel to the boundary and
$x_\perp$ is orthogonal to the boundary.  The one-point function in a BCFT
behaves kinematically  like a CFT two-point function
which is given  by \cite{Cardy:1984bb}
\begin{equation}
    \langle \mathcal{O}(x) \rangle = \frac{A_{\cal O}}{(2x_\perp)^\Delta} ,
\end{equation}
where the coefficient $A_{\cal O}$ is a physical parameter depending on both
the operator and the boundary condition, and  $\Delta$ is the scaling
dimension of the operator $\mathcal{O}$.
Similarly, the two-point function of scalar operators in
a BCFT behaves kinematically like a CFT four-point function
\begin{equation}\label{bcft-2pt}
  \langle\mathcal{O}(x)\mathcal{O}(y)\rangle =
  \frac{1}{|4x_\perp y_\perp|^\Delta} \mathcal{G}(\xi)\;,
\end{equation}
and depends on an
undetermined function of a single conformal-invariant cross-ratio
$\xi$ 
\begin{equation}
  \xi = \frac{(x - y)^2}{4x_\perp y_\perp} =
  \frac{\pm (x_0 - y_0)^2 + (\vec{x} - \vec{y})^2 + (x_\perp - y_\perp)^2}
       {4x_\perp y_\perp} .
    \label{eq:cross-ratio}
\end{equation}
Here the $+$ sign is for the Euclidean theory and the $-$ sign is for the
Lorentzian theory.
The cross ratio $\xi$ takes positive values when the two
operators live in the Euclidean signature or are space-like separated
in the Lorentzian signature. The two point correlation function in a
BCFT can be expanded in two different OPE limits namely the bulk limit
$\xi \to 0$ and the boundary limit $\xi\to \pm \infty$
\footnote{
In Euclidean signature, the boundary limit is given by $\xi \to \infty$.
In Lorentzian signature and for a time-like interval, it is also possible to
reach the boundary in the limit of $\xi \to -\infty$. 
}
as
\begin{equation}
  \mathcal{G}(\xi)=\sum_{\mathcal{O}}\lambda_{\mathcal{O}}g_{\Delta_\mathcal{O}}^B(\xi)
  =\sum_{\mathcal{\hat{O}}}\mu_{\mathcal{\hat{O}}}g_{\Delta_\mathcal{\hat{O}}}^b(\xi),
\end{equation}
where $g^B$ and $g^b$ are the bulk and boundary conformal blocks
respectively, and the sum is over the set of bulk primary operators
${\mathcal{O}}$ or boundary primary operators $\hat{\mathcal{O}}$
which appear in the corresponding OPEs.

The Euclidean BCFT correlation function have singularities only when
the operators approach each other or when they approach the boundary
(which can be thought as an operator approaching their mirrored double
across the boundary). When the two operators approach each other and
away from the boundary, the BCFT two-point function in the bulk limit
$\xi\rightarrow 0$ behaves like \cite{Reeves:2021sab}
\begin{equation}\label{bulk-ope}
  \langle\mathcal{O}(x)\mathcal{O}(y)\rangle= \frac{1}{|{x - y|}^{2\Delta} }
  + \ldots , \quad {\cal G}(\xi) \sim \xi^{-\Delta},\qquad
  \mbox{as $\xi \to 0$}.
\end{equation}
Similarly, the two point BCFT correlator in the boundary
limit $\xi \rightarrow \infty$ when the operators are much closer
to the boundary than to each other, the correlator behaves like
\cite{Reeves:2021sab}
\begin{equation}\label{boundary-ope}
  \langle\mathcal{O}(x)\mathcal{O}(y)\rangle=
  \frac{\mathcal{A}^2}{|{4 x_\perp y_\perp|}^{\Delta}} + \ldots ,
  \quad \mathcal{G}(\xi) \sim \mathcal{A}^2,
  \qquad \mbox{as $\xi \to \infty$}
\end{equation}
for some constant  $\mathcal{A}$. 
When the operators $\mathcal{O}(x)$ and $\mathcal{O}(y)$ are time-like
separated in the Lorentzian signature, the two point function can be
obtained by an analytic continuation of $\mathcal{G}(\xi)$ to the
time-like region $\xi<0$ around the branch-point at $\xi=0$. Then,
there exists another singularity in the limit $\xi = -1$, known as
Regge limit of the BCFT \cite{Mazac:2018biw} where $\mathcal{O}(y)$
approaches the light-cone of the mirror reflection of $\mathcal{O}(x)$
with the boundary behaving like the mirror. The BCFT two point
function diverges at the Regge limit $\xi \to -1$ as
\cite{Mazac:2018biw,Reeves:2021sab}
\begin{equation}\label{regge-limit}
  \langle\mathcal{O}(x)\mathcal{O}(y)\rangle=
  \frac{1}{|{4 x_\perp y_\perp|}^{\Delta}}\frac{1}{(1+\xi)^{\Delta} }+ \ldots,
  \quad \mathcal{G}(\xi) \sim (\xi+1)^{-\Delta}, 
  \qquad \mbox{as $\xi \to -1$}.
\end{equation}
As shown in \cite{Maldacena:2015iua}, a CFT correlation function have
singularities at the location of the
Landau diagrams where null
particles interact at local vertices. One can argue similarly that
\cite{Reeves:2021sab}
the only singularities of a
BCFT two-point function lie on the light-cone and its reflection.

\section{Time-like entanglement entropy in BCFT${}_2$}\label{TEE-bcft}
In this section, we compute  the time-like entanglement entropy
for a time-like interval at zero and finite
temperature in a two dimensional Lorentzian BCFT.
We will do so by first computing the
entanglement entropy for a generic interval in Euclidean signature
followed by an analytical continuation to the Lorentzian signature to
obtain the time-like entanglement entropy.

\subsection{Zero temperature}

\begin{figure}[ht]
	\centering
	\includegraphics[scale=1]{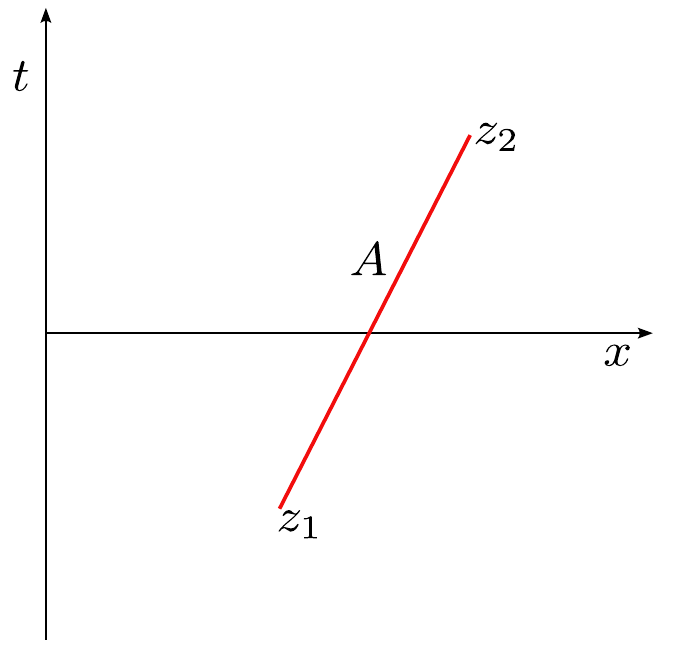}
	\caption{Schematic of a generic interval $A$ in two dimensional BCFT. }
	\label{TEE-interval}
\end{figure}

Consider the configuration depicted in \cref{TEE-interval} for 
an interval $A\equiv [z_1,z_2]:= 
[t_1,x_1),(t_2,x_2)]$ at zero temperature in a BCFT.
The entanglement entropy for a generic space-like interval $A$ in a
BCFT${}_2$ may be expressed in terms of twist field correlators using
the replica technique \cite{Calabrese:2004eu,Calabrese:2009qy} as
\begin{equation}\label{ent-replica}
S_A=\lim_{n\to 1}\frac{1}{1-n}\log\left<\sigma_n(z_1)\bar{\sigma}_n(z_2)\right>,
\end{equation}
where the scaling dimensions of the twist fields $\sigma_n$ and
$\bar{\sigma}_n$ are given by
\begin{equation}
\Delta_n=\bar{\Delta}_n=\frac{c}{12}\left(n-\frac{1}{n}\right).
\end{equation}
We note that for a purely time-like interval $A$ at a distance $x_1 = x_2$
from the boundary having time-like width $T_0= t_2 -t_1$,
the cross-ratio which is given by
\be
\xi = -\frac{T_0^2}{4 x_1^2}\,,
\ee
is always negative. It is instructive to illustrate the different limits
of $\xi$ by following the value of $T_0$ with respect to $x_1$.
When $T_0 \ll x_1$, we have $\xi \to 0^-$ and we are in the bulk limit.
As $T_0$ increase and reaches the region of $2x_1$,  one of the two points of
the interval approaches the light cone of the image of the other point
and we have the Regge limit $\xi \to -1$. 
Finally for $T_0 \gg x_1$,
we arrive at the boundary limit $\xi \to -\infty$. This is
different from the other boundary limit $\xi \to \infty$ which is obtained
for a space-like interval. Let us now
consider these limits in detail.

\subsection*{I. Bulk limit}

Consider first the case that the interval is far away from the boundary
such that $\xi \to 0$.
In this case, the
dominant contribution of the two point function
comes from the bulk channel OPE.
On utilizing the \cref{bulk-ope}, the two point twist in the
bulk limit may be written in the following form
\begin{equation}\label{2pt-bulk}
  \left<\sigma_n(z_1)\bar{\sigma}_n(z_2)\right>
  =\frac{\epsilon^{2\Delta_n}}{|z_1-z_2|^{2 \Delta_n}},
\end{equation}
where $\e$ is a UV regulator.
Now using the above \cref{2pt-bulk} in \cref{ent-replica}, the
entanglement entropy for a generic space-like interval $A$
in the Lorentzian signature is given by
\begin{equation}
S_A=\frac{c}{3}\log\frac{\sqrt{(x_1-x_2)^2-(t_1-t_2)^2}}{\epsilon}
\end{equation}
For a
purely time-like interval with $x_1=x_2$,
the time-like entanglement entropy $S_A^T$ in the bulk limit is given by
\begin{equation}\label{tee-bulk}
S_A^T=\frac{c}{3}\log\frac{T_0}{\epsilon}+\frac{i \pi c}{6} := S^B,
\end{equation}
where here the superscript $B$ stands for the bulk. 
We observe that the time-like entanglement entropy is complex in this
phase and resembles the usual CFT$_2$ result as given in
\cref{TEE-cft}. Note that the standard entanglement entropy in a
BCFT${}_2$ also resembles the entanglement entropy of an interval in a
CFT$_2$ in the bulk channel.

\subsection*{II. Regge limit}

\begin{figure}[ht]
	\centering
	\includegraphics[scale=1]{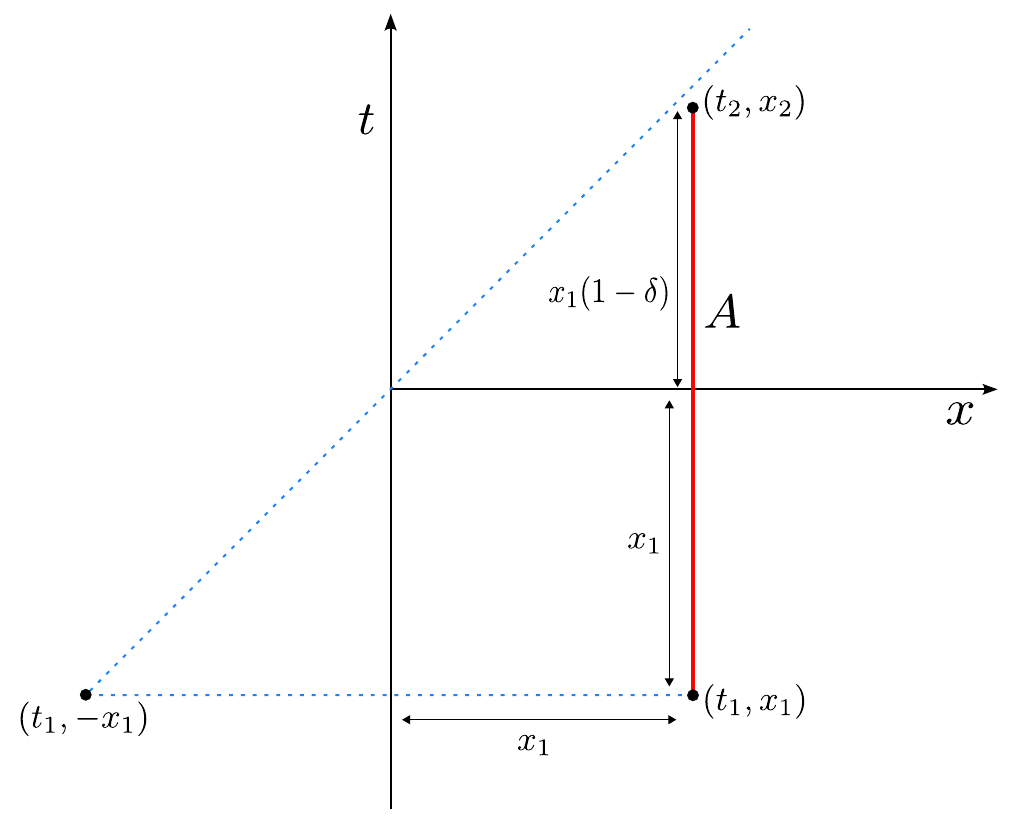}
	\caption{Schematic of a time-like interval in the Regge limit. }
	\label{Regge-phase}
\end{figure}

As $T_0$ increases, it will eventually reach the region $T_1 \approx 2 x_1$.
Let us parameterize the time interval by  $T_0 = 2x_1(1-\d /2)$. Due
to time translational invariance, we can take the purely time-like
interval as 
$A[(t_1,x_1),(t_2,x_2)] \equiv
\left[(-x_1,x_1),(x_1 (1-\d),x_1)\right]$. See \cref{Regge-phase}.
In the leading order of small $\delta$,
the cross ratio $\xi$ is given by
\begin{equation}
\xi=\frac{-(t_2-t_1)^2+(x_2-x_1)^2}{4x_1 x_2}=- (1-\d)
\end{equation}
and
the Regge limit is attained when the second end point
of the interval proximate the light cone of
the mirror image of the first end point of the interval.
Using the two point function
\begin{equation}\label{2pt-regge}
  \left<\sigma_n(z_1)\bar{\sigma}_n(z_2)\right>
  =\frac{1}{|2 x_1|^{2\Delta_n}}\frac{1}{(1+\xi)^{\Delta_n}}, 
\end{equation}
one obtains immediately the time-like entanglement entropy for the
time-interval
$A$ in the Regge limit as
\begin{equation}\label{tee-regge}
  S_A^T=
  \begin{cases}
    \frac{c}{3}\log\left(
  \frac{2 x_1}{\e} \sqrt{2 -\frac{T_0}{x_1}}\right), \quad
  \mbox{for $T_0 \to 2x_1^-$}, \quad \mbox{i.e.}\, \xi \to -1^+,\\
  \frac{c}{3}\log\left(
  \frac{2 x_1}{\e} \sqrt{\frac{T_0}{x_1}-2}\right) + \frac{i \pi c}{6}, \quad
  \mbox{for $T_0 \to 2x_1^+$}, \quad \mbox{i.e.}\, \xi \to -1^-,
  \end{cases}
  := S^R,
\end{equation}
where here the superscript $R$ stands for Regge.
Note that $\xi =-1$ is a branch point and this explain why
$S^R$ picks up an imaginary part as $T_0$ crosses the value of
$2x_1$. Physically, the point $z_2$ has gone from the inside 
of the light-cone 
of the image point $z_1'$ to the outside.

\subsection*{III. Boundary limit}

In the boundary limit, the two point twist correlator may be
expressed as
\begin{equation}\label{2pt-bdy}
  \left<\sigma_n(z_1)\bar{\sigma}_n(z_2)\right>=\frac{g_b^{2(1-n)}
    \epsilon^{2\Delta_n}}{|4 x_1 x_2|^{\Delta_n}},
\end{equation}
where $\mathcal{A}=g_b^{(1-n)}$ arises from the replica technique for
BCFT entanglement entropy computations \cite{Sully:2020pza}. This gives the
entanglement entropy for a generic interval $A$ as
\begin{equation}\label{gen-ent-bdy}
  S_A=\frac{c}{6}\log\frac{2 x_1}{\epsilon}+\frac{c}{6}\log\frac{2 x_2}
  {\epsilon}+2\log g_b,
\end{equation}  
where the second term describes the boundary entropy which depends on the boundary condition. Note that \eq{gen-ent-bdy}
is independent of the time $t$ coordinate. Now taking $x_1=x_2$
for a pure time-like interval, we obtain the time-like entanglement entropy 
\begin{equation}\label{tee-bdy}
S_A^T=\frac{c}{3}\log\frac{2 x_1}{\epsilon}+2\log g_b := S^b,
\end{equation}
where here the superscript $b$ stands for the boundary.
We see that the time-like entanglement entropy is real in this phase and
includes boundary entropy which is expected in the boundary phase.

\begin{figure}[ht]
	\centering
	\includegraphics[scale=0.4]{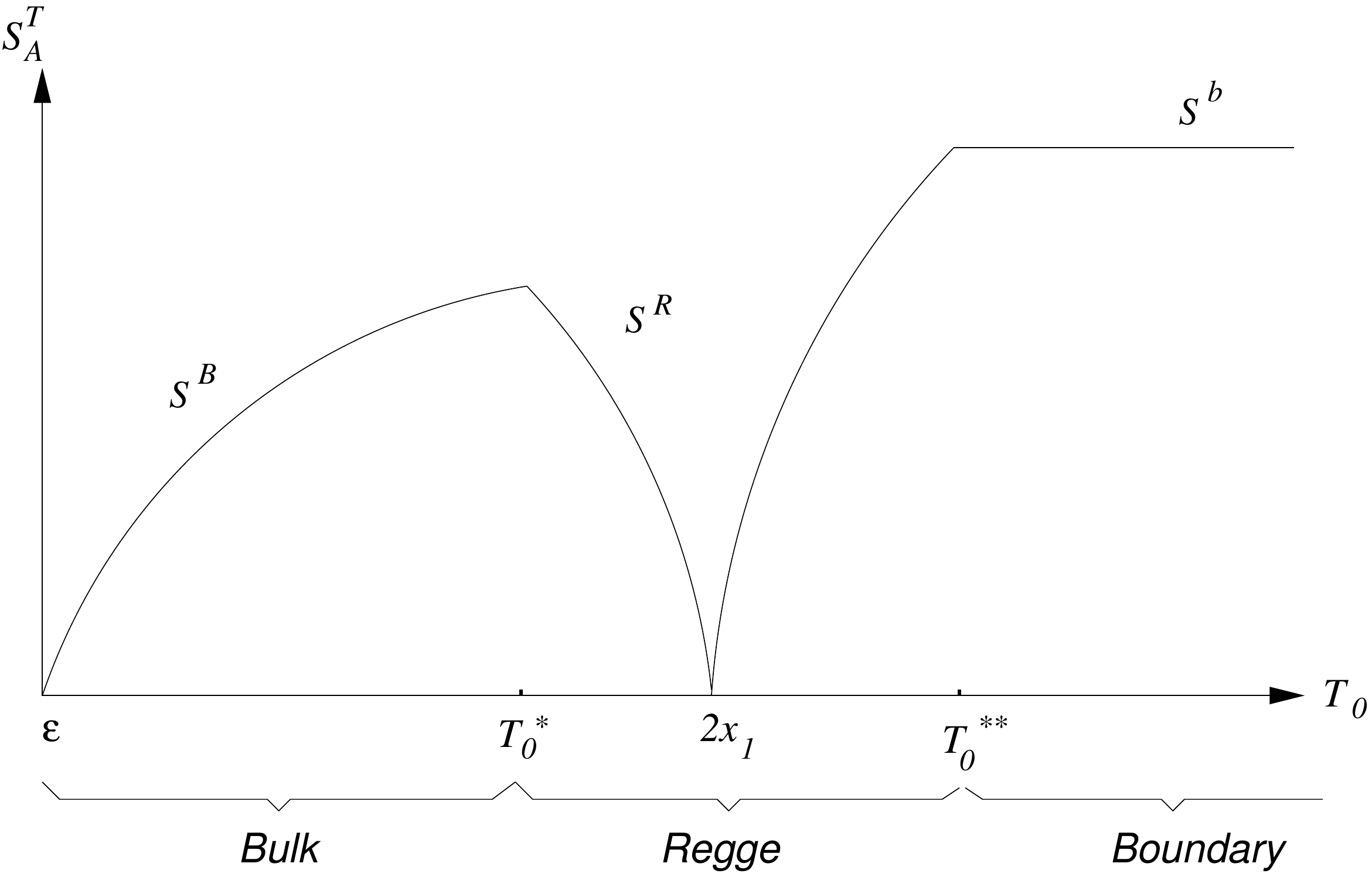}
	\caption{Phases of the time-like entanglement entropy $S^T_A$ in BCFT.}
	\label{curve}
\end{figure}

A couple of remarks are in order.
\ben
\item It is interesting that in contrast to the usual two phases of standard
entanglement entropy in BCFT$_2$, we have a new phase of time-like
entanglement entropy \eq{tee-regge}
which arises from the light-cone singularities. This
is possible only for a time-like interval.
\item
We remark that as the time-like entanglement entropy $S^T_A$
is obtained by taking the
log of the 2-point function \eq{bcft-2pt} and 
$\cG$ behaves as $\cG \sim \xi^{-\D_n}, (1+\xi)^{-\D_n}$, {\rm constant} $\cA^2$
in the above
said limits, it is clear that $S^T_A \sim \log \cG$
picks up a non vanishing
imaginary part in the bulk limit $\xi \to 0^-$ and in the time-like side
of the Regge limit $1+\xi \to 0^-$, i.e $ T_0 \gtrapprox 2x_1$.
\item It is instructive to follow the behavior of the time-like entanglement
entropy and discuss its phase transition. Without loss of generality,
consider fixed $x_1$ and let $T_0$
changes. For small $T_0$, $S^T_A$ is given by $S^B$ of \eq{tee-bulk}.
As $T_0$ increases, cross over from the bulk to the Regge behaviour occurs at:
\be
S^B = S^R: \quad T_0 = 2(\sqrt{3}-1) x_1 := T_0^*,
\ee
where the equality of entropies is for the real part.
As $T_0$ continue to increase, it eventually reaches the light-cone
singularity point $T_0 \approx 2 x_1$ and passes it. Crossover from the Regge
behaviour to the boundary behaviour occurs at:
\be
S^R = S^b: \quad T_0 = (2+ g_b^{12/c}) x_1 := T_0^{**}.
\ee
One can show that $S_{bdy} \geq 0$ in an unitary BCFT and so
$T_0^{*} <2x_1 \leq T_0^{**}$.
The behaviour of the time-like entropy curves is shown in  \cref{curve}.

\een

\subsection{Finite temperature}
Next, let us compute the time-like entanglement entropy of an interval
in two dimensional CFT on the half line at finite temperature
$\beta^{-1}$. This can be obtained by computing the two point twist
correlation function on a half cylinder (with $w=x+i\tau, x\leq 0$)
where the Euclidean time direction $\tau$ is compactified with period
$\beta$. It is convenient to express it in terms of the
two  point function on the interior of the unit disc $z$ via the 
conformal transformation $ w \to z = e^{\frac{2\pi w}{\beta}}$:
\begin{equation}\label{2-pt-transf}
\left<\sigma_n(w_1, \bar{w}_1)\bar{\sigma}_n(w_2,\bar{w}_2)\right>_{\rm cyl}
 =  \prod_{i=1}^{2}
 \left ( \frac{ dz }{ d w } \right )^{  \Delta_n/2 }_{w=w_i}
 \left ( \frac{ d \bar{z} }{ d \bar{w} }
 \right )^{\bar{\Delta}_n/2 }_{\bar{w}=\bar{w}_i}
 \left<\sigma_n(z_1,\bar{z}_1)\bar{\sigma}_n(z_2,\bar{z}_2)\right>_{\rm disc},
\end{equation}
where the two point function
on the unit disk\footnote{The correlator on the unit disk can be obtained
from the correlator on the upper half plane (UHP) through the M\"{o}bius
transformation as $u=-i\left(\frac{z-1}{z+1}\right)$.}
of primary operators of dimensions $\Delta_1$ and $\Delta_2$ is given by
\cite{Calabrese:2009qy,Calabrese:2014yza}
\begin{equation}\label{2-pt-disk}
  \left<\sigma_1(z_1,\bar{z}_1)\bar{\sigma}_2(z_2,\bar{z}_2)\right>_{\rm disc}
  =\frac{1}{(1-z_1\bar{z}_1)^{\Delta_1}(1-z_2 \bar{z}_2)^{\Delta_2}}
  \mathcal{G}({\xi}).
  \ee
Here $\xi$ is the cross ratio
\be
  \xi=\frac{|z_1-z_2| |\bar{z}_1-\bar{z}_2|}{(1-z_1\bar{z}_1)(1-z_2\bar{z}_2)}.
  \ee
  and the function $\mathcal{G}({\xi})$ is known explicitly only in
  the bulk, boundary or Regge limit. The entanglement entropy for a generic
  interval $A = [w_1, w_2]$ on the cylinder is given by
  \be
  S_A = \lim_{n \to 1} \frac{1}{1-n}\log\left<\sigma_n(w_1,\wb_1)
  \bar{\sigma}_n(w_2,\wb_2)
  \right>_{\rm cyl}.
  \ee
As a result, we have
  three phases for the TEE at finite temperature   similar to the zero
  temperature case.

\subsection*{I. Bulk limit}

For this case when the interval is far away from the boundary, we have
$\cG (\xi) \sim \xi^{-\D_n}$ and hence
\be
S_A
=\frac{c}{6}\log\left[ \left(\frac{\beta}{\pi\epsilon}\right)^2
  \sinh\frac{\pi w_{12}}{\beta}\sinh\frac{\pi \bar{w}_{12}}{\beta}\right],
\ee
where $w_{12}=-x_{12}+i\tau_{12}$ and $\bar{w}_{12}=-x_{12}-i
\tau_{12}$ are the interval length of $A$ on the cylinder. Now
analytically continue to the Lorentzian signature via $\tau\to i t$, we have
\be
S_A = \frac{c}{6} \log \left[ \frac{1}{2}
  \left(\frac{\beta}{\pi\epsilon}\right)^2
(\cosh \frac{2\pi x_{12}}{\b} - \cosh \frac{2\pi t_{12}}{\b}) 
  \right].
\ee
Setting $x_{12}=0$ and $t_{12}=T_0$, one obtain the
time-like entanglement entropy for an interval $A$
at finite temperature 
\begin{equation}\label{fin-t-bcft-I}
  S_A^T=\frac{c}{3}\log\left[\frac{\beta}{\pi\epsilon}
    \sinh\left(\frac{\pi T_0}{\beta}\right)\right]+\frac{i \pi c}{6}.
\end{equation}
We observe that the above expression is same as
the CFT result \cref{tee-finte-temp-cft}
which is expected because the interval is far away from the boundary
in this phase.

\subsection*{II. Regge limit}
In the Regge limit, we have $\cG \sim (1+\xi)^{-\D_n}$ and
the finite temperature entanglement entropy for
a generic interval in the Regge limit 
is given by
\begin{equation}
  S_A=\frac{c}{6}\log\left[\left( \frac{\beta}{\pi \epsilon}\right)^2
\left(
\sinh \frac{\pi (w_1+\bar{w}_1)}{\beta}
\sinh \frac{\pi (w_2+\bar{w}_2)}{\beta}+
  \sinh\frac{\pi w_{12}}{\beta}\sinh\frac{\pi \bar{w}_{12}}{\beta}
  \right)
  \right].
\end{equation}
Analytically continuing to the Lorentzian signature via $\t
\to it$, we get the following time-like entanglement entropy of a generic time-like
interval as
\begin{equation}
  S_A=\frac{c}{6}\log\left[\left( \frac{\beta}{\pi \epsilon}\right)^2
    \left(\sinh \frac{2\pi x_1}{\beta}
    \sinh \frac{2\pi x_2}{\beta} -
    \sinh\frac{\pi (x_{12}+t_{12})}{\beta}
    \sinh\frac{\pi (-x_{12}+t_{12})}{\beta}
    \right)
    \right].
\end{equation}
For a purely time like interval $T_0$ placed at a distance $x_1$ from the
boundary, a light-cone singularity is reached 
when the second end point of the
interval $A$ approaches the mirror reflection of light cone of first
point of the interval. Parameterize $T_0 = 2x_1 (1-\d/2)$ as before,
the time-like entanglement entropy for a time interval
at finite temperature in the Regge limit is given by
\begin{equation}\label{fin-t-bcft-III}
  S_A^T=
  \begin{cases}
    \frac{c}{6}\log\left(
  \frac{\beta (2x_1 -T_0)}{\pi \e^2} 
  \sinh \frac{4\pi x_1}{\beta}\right),
   \quad
   \mbox{for $T_0 \to 2x_1^-$}, \quad \mbox{i.e.}\, \xi \to -1^+,\\
    \frac{c}{6}\log\left(
  \frac{\beta (T_0- 2x_1)}{\pi \e^2} 
  \sinh \frac{4\pi x_1}{\beta}\right) + \frac{i \pi c}{6}, \quad
\mbox{for $T_0 \to 2x_1^+$}, \quad \mbox{i.e.}\, \xi \to -1^-,
  \end{cases}
  := S^R.
\end{equation}
Note that this
has the correct zero temperature limit \eq{tee-regge} as expected.
Note also that the discontinuity of the imaginary part at the branch point 
is the same as in the zero temperature case.

\subsection*{III. Boundary limit}

As the interval get close to the boundary, we have
$\mathcal{G}({\xi})=g_b^{2(1-n)}$ in the UHP for the boundary limit and one
obtains the entanglement entropy for a generic interval at
finite temperature as
\begin{equation}
\begin{aligned}
  S_A=\frac{c}{6}\log\left[\left( \frac{\beta}{\pi \epsilon}\right)^2\sinh
    \frac{\pi (w_1+\bar{w}_1)}{\beta}\sinh \frac{\pi (w_2+\bar{w}_2)}{\beta}
    \right]+2 \log g_b.
\end{aligned}
\end{equation}
Analytically continuing to the Lorentzian signature and take
$w_1+\bar{w}_1=-2x_1$, $w_2+\bar{w}_2=-2 x_2$ in the above expression we get
\begin{equation}\label{finite-t-gen-exp}
\begin{aligned}
  S_A=\frac{c}{6}\log\left[\left( \frac{\beta}{\pi \epsilon}\right)^2\sinh
    \frac{2\pi x_1}{\beta}\sinh \frac{2\pi x_2}{\beta}\right]+2 \log g_b.
\end{aligned}
\end{equation}
Note  that this is independent of the time coordinates.
The time-like entanglement entropy for a purely time-like interval
at finite temperature in this phase may now be obtained by taking $x_1=x_2$
in \cref{finite-t-gen-exp} and we obtain
\begin{equation}\label{fin-t-bcft-II}
\begin{aligned}
  S_A^T=\frac{c}{3}\log\left( \frac{\beta}{\pi \epsilon}\sinh
  \frac{2\pi x_1}{\beta}\right)+2 \log g_b.
\end{aligned}
\end{equation}
We observe that the time-like entanglement entropy
is real in this phase unlike the previous phase.

One can similarly discuss the crossover of the entropies.
The phase diagram is similar
to the zero temperature case qualitatively except that the crossover position is
now temperature dependent. We will omit the details here.

\section{Holographic Time-like entanglement entropy in AdS${}_3$/BCFT${}_2$ }
\label{HTEE}
We now obtain the holographic time-like entanglement entropy in the
context of AdS${}_3$/BCFT${}_2$ which involves the computation of bulk
extremal curves (geodesics). We further show that the holographic TEE
matches exactly with the dual field theory results.

\subsection{AdS/BCFT duality}

BCFTs \cite{Cardy:2004hm,McAvity:1993ue} describe physical 
systems with boundaries at the critical point.  In addition to the
traditional field theory techniques,
a novel non-perturbative AdS/BCFT correspondence
was originally introduced by Takayanagi \cite{Takayanagi:2011zk}
based on the idea of holography. 
At the level of classical gravity, the action for AdS/BCFT is given by
\begin{equation}\label{action}
  I=\frac{1}{16\pi G_N}
  \int_{N}d^{d+1}x \sqrt{|g|}( R-2\Lambda )
  +\frac{1}{8\pi G_N}\int_Q d^dy\sqrt{|h|} (K -T),
\end{equation}
where $K$ is the extrinsic curvature, $T$ is the tension of
end-of-the-world (EOW) brane $Q$ and $h_{ij}$ is the induced metric on $Q$.
The original proposal of  Takayanagi
\cite{Takayanagi:2011zk} is to take
on $Q$ the Neumann boundary condition (NBC) 
\begin{equation}\label{NBC}
\text{NBC}: \ K^{ij} -(K-T)h^{ij} =0.
\end{equation}
The NBC 
imposes conditions on the end-of-the-world
brane $Q$ \cite{Takayanagi:2011zk,Fujita:2011fp,Nozaki:2012qd}
as well as the bulk Einstein metric \cite{Miao:2017aba}.
In addition to this, one may impose alternative boundary conditions
for AdS/BCFT such as the conformal boundary condition (CBC)
\cite{Miao:2017gyt,Chu:2017aab} 
which fixes the conformal geometry
and the trace of the extrinsic curvature of $Q$
\be \label{CBC}
\text{CBC}: \ K=\frac{d}{d-1} T.
\ee
One may also impose  the
Dirichlet boundary condition (DBC) \cite{Miao:2018qkc}, all of which
define consistent theory of AdS/BCFT. For the special case of BCFT$_d$ on a
half space, say $x \geq 0$,
the bulk geometry is given by a portion of AdS$_{d+1}$.
In the Poincar\'{e} coordinates (take AdS radius to be 1 for convenience)
\be
ds^2=\frac{1}{z^2}(dz^2+dx_i^{2})\ ,
\ee
where the bulk geometry is given by a wedge with the EOW brane
located at $x= z \sinh \r_0 $. Here $\r_0$ is determined in term of the
``tension''
$T$ as $T = (d-1) \tanh \r_0$. Note that for flat boundary, all three
BC give the same EOW brane \cite{Miao:2017gyt,Chu:2017aab,Miao:2018qkc}.
However the holographic BCFT are still different in general as the spectrum of graviton fluctuations on the EOW brane is different
for the different choice of BC. This, for example, results in different
two point functions for the respective holographic BCFTs
\cite{Chu:2021mvq}. The existence of massive gravitons also leads to some
puzzling
behaviour for the island proposal as 
was pointed out originally in \cite{Geng:2020fxl,Geng:2021hlu}.

Having reviewed about the AdS/BCFT duality, we now turn our attention
to the computation of holographic time-like entanglement entropy for a
time-like interval at zero and finite temperature in holographic 
BCFT${}_2$.

\subsection{Zero temperature}
Consider a pure
time-like interval $A$ in the dual BCFT${}_2$ at a fixed distance
$x=x_1$ from the boundary. We have three possible choices of RT
surface for this configuration depending on the size and distance from
the boundary of an interval which are described below.

\subsection*{Phase I: Bulk phase}
In this phase, the interval is far away from the boundary such that the
RT surface consists of two space-like geodesics and one time-like geodesic
similar to \cref{TEE-cft-review}. This configuration is depicted in
\cref{HTEE-I}. 
\begin{figure}[ht]
	\centering
	\includegraphics[scale=0.6]{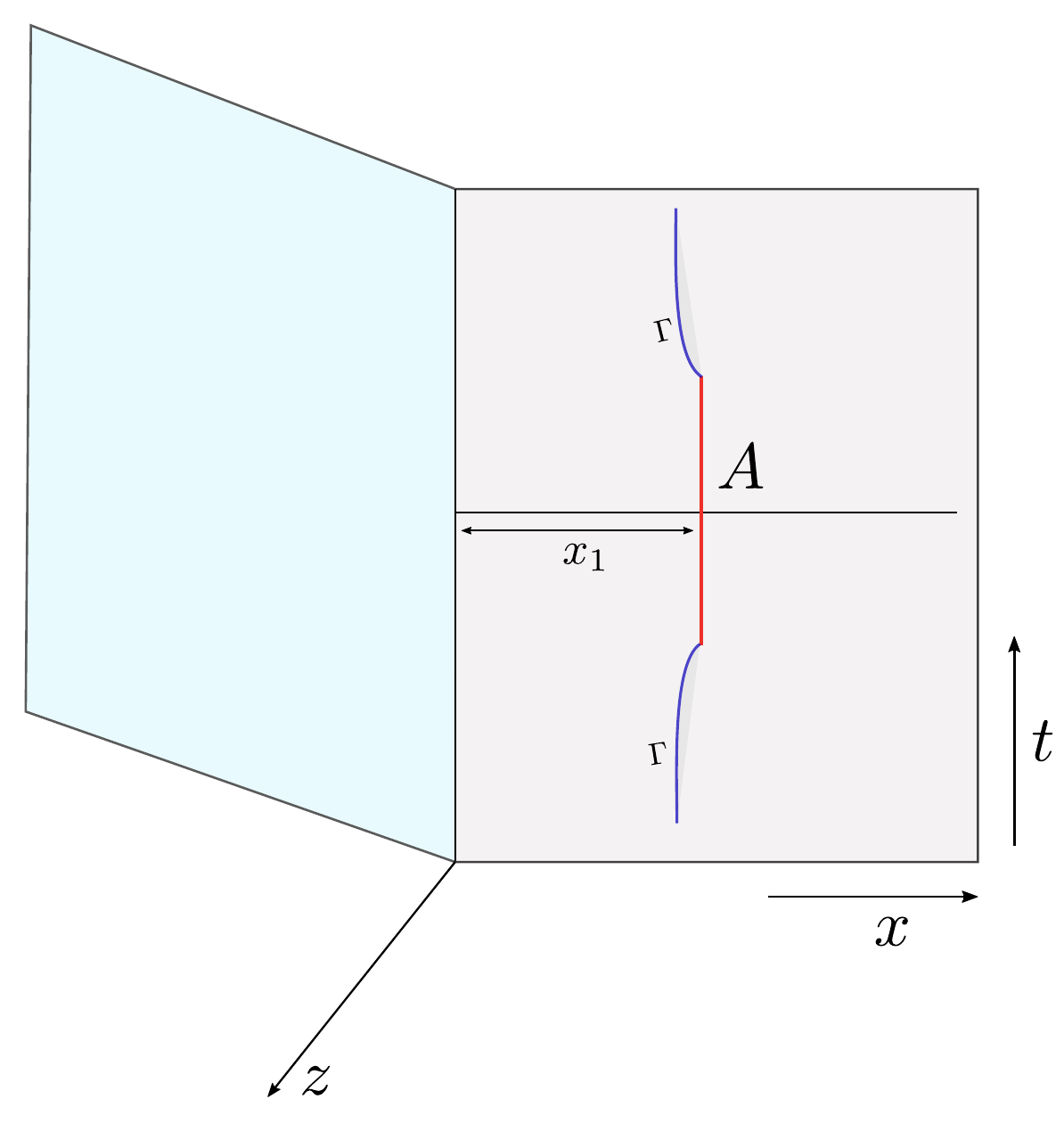}
	\caption{RT surfaces for configuration of a time-like interval
          in the bulk phase. }
	\label{HTEE-I}
\end{figure}
So the holographic time-like entanglement entropy for an interval
having length $T_0$ along the time direction in this phase may be expressed as
\begin{equation}
S_A^T=\frac{c}{3}\log\frac{T_0}{\epsilon}+\frac{i \pi c}{6},
\end{equation}
where the real part comes from the length of two hyperbola going to
future infinity $(t=+\infty, z=+\infty)$ and past infinity
$(t=-\infty, z=+\infty)$; and the imaginary part comes from the
geodesic joining the future and past infinities.
The above result
agrees exactly with the corresponding dual BCFT${}_2$ result
\cref{tee-bulk}.

\subsection*{Phase II: Regge phase}

For this phase,  the end points of the
purely time-like interval on the boundary can be parametrized as
$A[(t_1,x_1),(t_2,x_2)]\equiv \left[(-x_1,x_1),(x_1 (1-\d),x_1)\right]$
with $\d = 2- T_0/x_1$.
Consider first the case of $T_0 <2x_1$ with $\d>0$. In this case,
the RT surface is given by two geodesics
where each geodesic joins the one end point of the interval and
ends on the plane perpendicular to the boundary as shown in \cref{HTEE-III}. 
\begin{figure}[ht]
	\centering
	\includegraphics[scale=0.6]{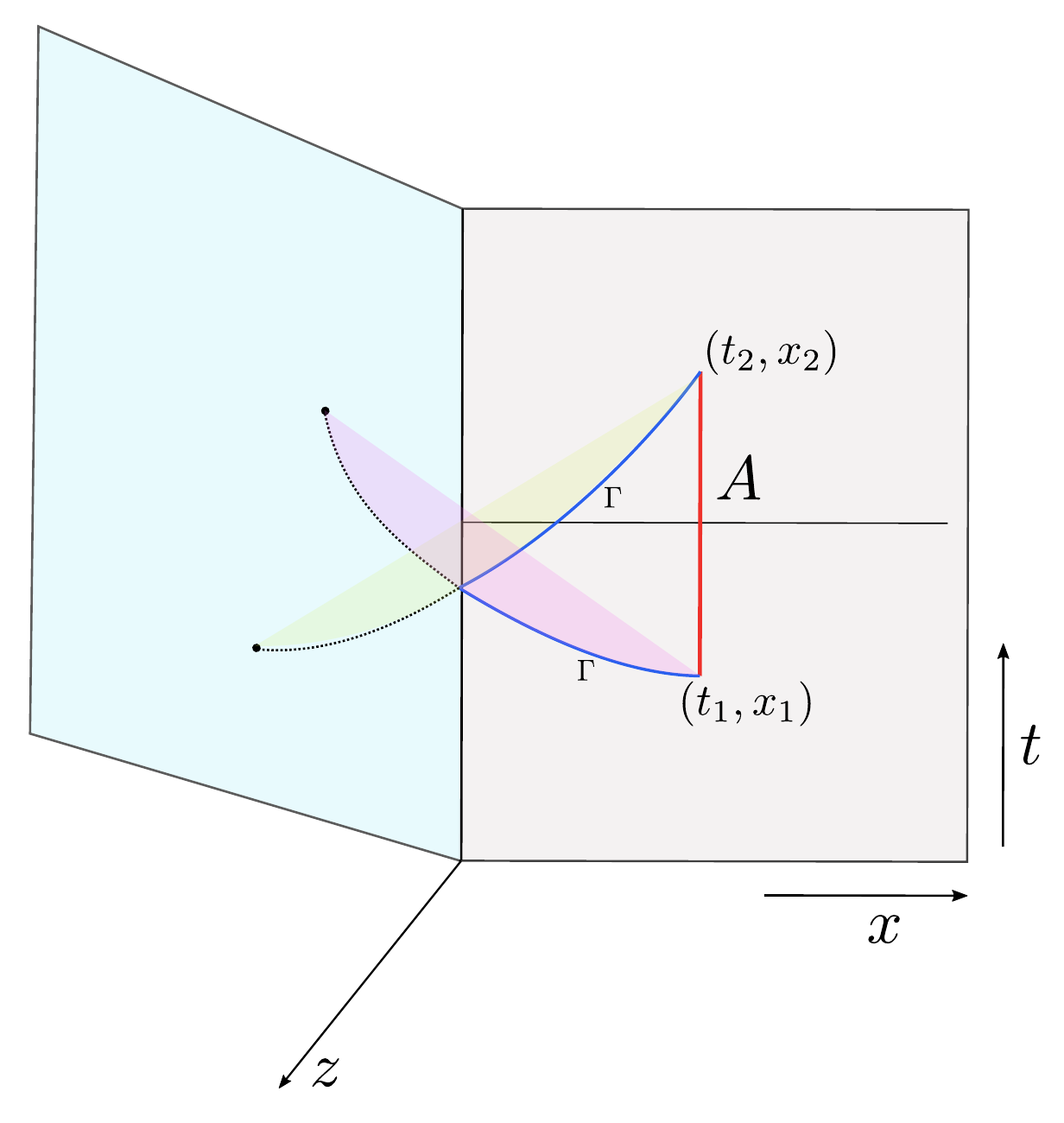}
	\caption{RT surfaces for configuration of a time-like interval
          in the Regge phase. }
	\label{HTEE-III}
\end{figure}
The geodesics lie along the $t=\pm mx+c$ plane.
The
plane passing through the points $(t_2,x_2)$ and mirror image of
$(t_1,x_1)$, i.e $(t_1,-x_1)$, is described by
\begin{equation}
t=mx+c, \hspace{4mm} m=1-\frac{\d}{2},
\end{equation}
where $m$ is the slope and $c$ is some constant.
The induced metric on this plane can be written as
\begin{equation}
\begin{aligned}
ds^2&=\frac{-dt^2+dz^2+dx^2}{z^2}
=\frac{(1-m^2)dx^2+dz^2}{z^2}\\
&=\frac{dy^2+dz^2}{z^2},
\end{aligned}
\end{equation}
where we have defined $y=\sqrt{1-m^2}x$. Observe that the above metric
in $(y,z)$ coordinate looks like a time slice of AdS$_3$ geometry in
Poincar\'{e} coordinates. So, the length of the geodesic $\Gamma$ from
the point $(y_1,z_1)$ to a point on the perpendicular plane
$(y_2,z_2)$ (which is located at $x=0$) along this plane is given
by\footnote{The geodesic distance between two points $(t_1,x_1,z_1)$
and $(t_2,x_2,z_2)$ in the Poincar\'{e} patch of AdS$_3$ is given by
the standard formula
\begin{equation}
L=\cosh^{-1}\left[\frac{-(t_2-t_1)^2+(x_2-x_1)^2+z_1^2+z_2^2}{2z_1z_2}\right].
\end{equation}  }
\begin{equation}\label{L-regge}
\begin{aligned}
L&=\cosh^{-1}\left[\frac{(y_2-y_1)^2+z_1^2+z_2^2}{2z_1z_2}\right]\\
&\approx \log\left[\frac{\d x_1^2+z^2}{\epsilon z}\right],
\end{aligned}
\end{equation}
where the point on the boundary and perpendicular plane is given by
$(y_1,z_1)=(\sqrt{\delta} x_1,\epsilon)$ and $(y_2,z_2)=(0,z)$
respectively. After extremization of $L$ with respect to z, we get the
length of the geodesic as follows
\begin{equation}
L=\log \frac{2 x_1 \sqrt{\d}}{\epsilon}.
\end{equation}
The length of other geodesic which lie along the plane passing through
the points $(t_1,x_1)$ and $(t_2,-x_2)$ can be computed similarly and
have the same length as $L$. So, we now obtain the holographic
time-like entanglement entropy which is given by the sum of these two
geodesics after using the RT formula as
\begin{equation}\label{hol-TEE-zer0--III}
  S_A^T=\frac{c}{3}\log \left(\frac{2 x_1}{\epsilon}\sqrt{2-\frac{T_0}{x_1}}
  \right).
\end{equation}
This agrees precisely with the corresponding dual field theory result
\cref{tee-regge}.
Similarly one can discuss the Regge limit from the other side $T_0 >2x_1$
with $\d<0$.
The computation is the same as above except now we have to take the negative root of the inverse $\cosh$ in \cref{L-regge} for a time-like geodesic. As a result, we obtain 
the corresponding dual theory result \eq{tee-regge} with
the constant imaginary part.
We remark that
in field theory, the Regge phase of the TEE arises from the singular
behaviour of the twist operator two point function in the Regge limit.
On the gravity side, this arises from a particular RT geodesic which becomes
null in the limit.

\subsection*{Phase III: Boundary phase}

Finally we consider the boundary phase where the interval is closer
to the boundary. In this phase,
the RT surface end on the brane as shown in \cref{HTEE-II}. 
\begin{figure}[ht]
	\centering
	\includegraphics[scale=0.6]{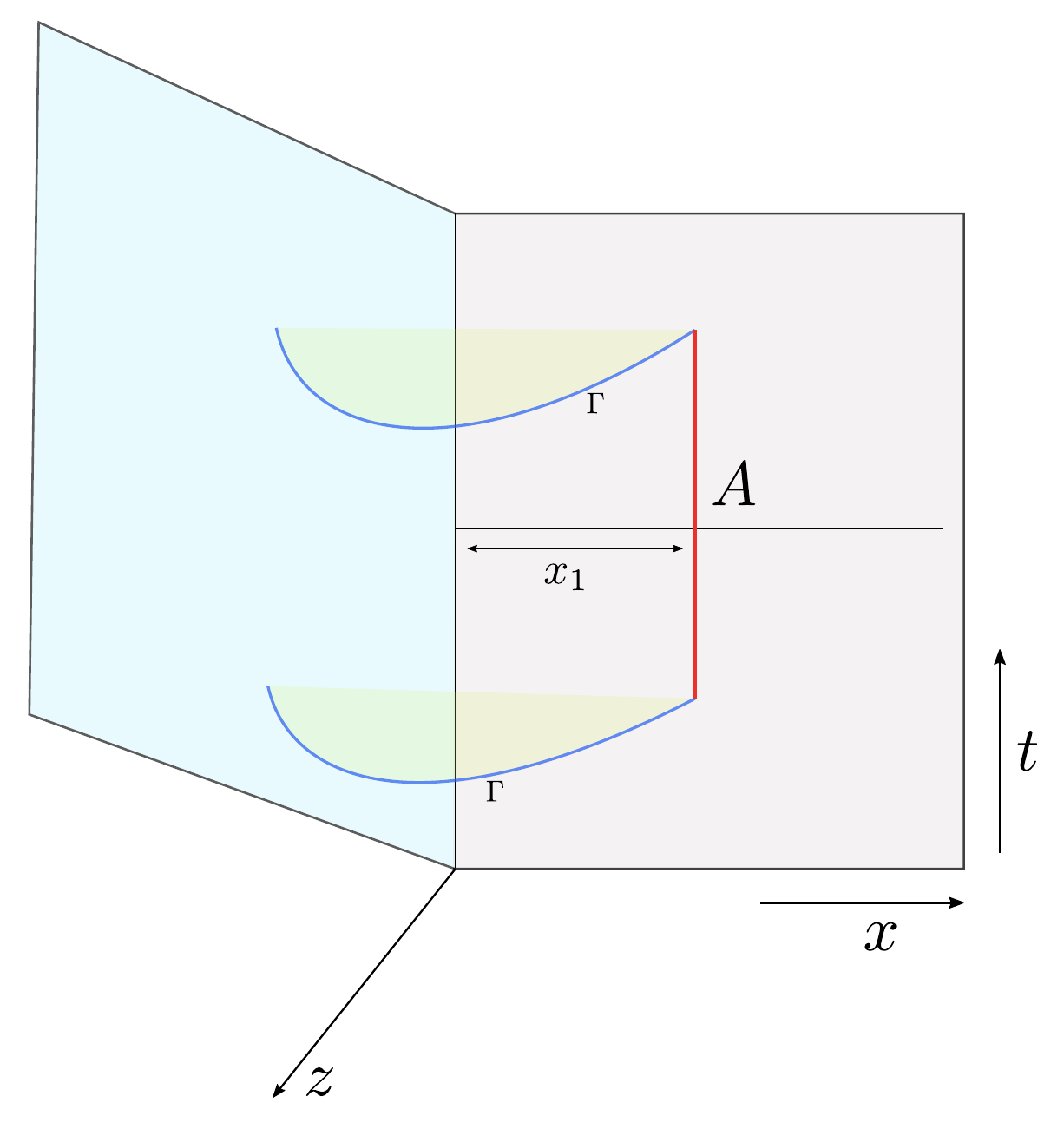}
	\caption{RT surfaces for configuration of a time-like interval in
         the boundary phase. }
	\label{HTEE-II}
\end{figure}
The length of each geodesic $\Gamma$ which is at a constant time slice is
given by \cite{Takayanagi:2011zk}
\begin{equation}\label{length-brane}
  L=\log\frac{2x_1}{\epsilon}+\rho_0,
\end{equation}
where $\rho_0=\tanh^{-1} (T )$. So the holographic
time-like entanglement entropy for a time-like interval $A$ may
obtained using the RT formula as
\begin{equation}
\begin{aligned}
S_A^T=\frac{c}{3}\log\frac{2x_1}{\epsilon}+ \frac{c}{3}\rho_0,
\end{aligned}
\end{equation}
where the second term describes the boundary entropy
\cite{Takayanagi:2011zk}. It is $T\geq 0$ for unitary theory and so
$S_{bdy} \geq 0$.
We observe that the holographic time-like
entanglement entropy in this phase also matches with dual BCFT result as described in
\cref{tee-bdy}.

It is interesting to note that the topology of the RT surfaces are
different for each of the three cases. This prompts us to interpret the crossover
behaviour of entropies as phase transition of the entropies.
Note that the bulk and boundary phases for the holographic
time-like entropy is similar to the usual holographic entanglement
entropy in AdS${}_3$/BCFT${}_2$. The bulk phase corresponds to the
connected RT surface while the boundary phase corresponds to the
disconnected RT surface. The appearance of the Regge phase is unique to the time-like entanglement entropy and is related
to the existence of geodesics lying along the plane that joins
one end point and the
mirror reflection of other end point of the interval.

\subsection{Finite temperature}
We consider BCFT${}_2$ in a half line at a finite temperature.
In this case, the only known dual bulk solution is for $S_{bdy}=T=0$ which is the BTZ black
hole as described in \cref{btz-metric} truncated by a tensionless EOW
brane perpendicular to the boundary along the $x=0$ direction\footnote{We note
that for a general non-zero $T$, the holographic dual of a BCFT on an interval at finite
temperature can be constructed by a part of two kind of bulk geometries i.e
thermal AdS$_3$ and the BTZ black hole \cite{Takayanagi:2011zk}. There
exist a Hawking-Page transition at a certain temperature between these
two geometries which depends on the brane tension. It would be interesting to obtain
the TEE in this kind of setting which we leave to the future.}. For a finite
temperature BCFT on a half line with vanishing boundary entropy, we
have three phase of the RT surface for the holographic time-like
entropy which is similar to the zero temperature case, as follows.

\subsection*{Phase I: Bulk phase}
In this phase, the RT surface consists of two space-like and one
time-like extremal surface in a BTZ geometry similar to the CFT case
as described in \cref{TEE-cft-review}. So, the holographic time-like
entanglement entropy for a time-like interval of length $T_0$ in this
phase is given by
\begin{equation}
  S_A^T=\frac{c}{3}\log\left[\frac{\beta}{\pi\epsilon}
    \sinh\left(\frac{\pi}{\beta}T_0\right)\right]+\frac{i \pi c}{6},
\end{equation}
where the imaginary contribution comes from the time-like
geodesic. The above TEE matches with the corresponding dual field
theory result in \cref{fin-t-bcft-I}.

\subsection*{Phase II: Regge phase}

For this phase, we have two RT surfaces with each geodesic connecting
the end point of the interval and end on the perpendicular EOW brane,
lying along the plane $t=mx+c$ similar to corresponding zero
temperature case. Since the EOW brane is perpendicular to boundary,
the length of each geodesic is half of the geodesic connecting the
point $(t_2,r_2,x_2)$ and mirror image of $(t_1,r_1,x_1)$ which is
$(t_1,r_1,-x_1)$. The geodesic length between two the points
$(t_1,r_1,x_1)$ and $(t_2,r_2,x_2)$ in the BTZ black hole geometry can
be expressed as
\begin{equation}\label{geod-btz-l}
  L=\cosh^{-1}\left[\frac{r_1 r_2}{r_{+}^2}\cosh r_+ (x_2-x_1)-
    \sqrt{\left(\frac{r_1^2}{r_+^2}-1\right)}
    \sqrt{\left(\frac{r_2^2}{r_+^2}-1\right)}\, \cosh r_+(t_2-t_1)\right].
\end{equation}

The end point of the time-like interval $A$ for this phase is given by
$A[(t_1,r_1,x_1),(t_2,r_2,x_2)]\equiv
[(-x_1,r_\infty,x_1),(x_1(1-\d), r_\infty, x_1)]$ with $T_0=2x_1 (1-\d/2)$ for
this configuration. Consider first the case $\d>0$,
the length of geodesic connecting
$(t_2,r_2,x_2)=(x_1 (1-\d), r_\infty,x_1)$ and
$(t_1,r_1,-x_1)=(-x_1,r_\infty,- x_1)$ can be obtained using the
\cref{geod-btz-l} in the limit $\delta\to 0$ as
\begin{equation}\label{L-Regge_T}
\begin{aligned}
  L&=\cosh^{-1}\left[\left(\frac{\beta r_\infty}{2 \pi}\right)^2
    \cosh\frac{2\pi}{\beta}2x_1
    -\left(\left(\frac{\beta r_\infty}{2 \pi}\right)^2-1\right)
    \cosh\frac{2\pi}{\beta}(2x_1 (1-\d/2))\right]\\
  &= \log\left(\frac{\b(2x_1-T_0)}{\pi \e^2} \sinh
  \frac{4\pi x_1}{\beta}\right),
\end{aligned}
\end{equation}
where $r_\infty>>1$ is related to the UV cut off as
$r_\infty=1/\epsilon$. Similarly, length of the geodesic connecting
$(t_1,r_1,x_1)$ and $(t_2,r_2,-x_2)$ gives the same (\ref{L-Regge_T}). So the holographic time-like entanglement
entropy for this phase may be obtained using the RT formula as
$S_A^T=\frac{1}{4G_N}(\frac{L}{2}+\frac{L}{2})$ and
\be
  S_A^T
  =\frac{c}{6} \log\left(\frac{\b(2x_1-T_0)}{\pi \e^2} \sinh
  \frac{4\pi x_1}{\beta}\right).
\ee
This result agrees exactly with the corresponding BCFT${}_2$
result obtained in \cref{fin-t-bcft-III}. Similarly, one can consider
approaching the Regge limit from the side of $T_0> 2x_1$ with $\d<0$.
The computation is the same as above and we obtain 
the corresponding dual theory result \eq{fin-t-bcft-III} with
the constant imaginary part.

\subsection*{Phase III: Boundary phase}
Similar to the above discussion,
we have two RT surfaces ending on the perpendicular EOW brane in phase
II. The length of this geodesic connecting the interval and brane in a
BTZ geometry can be obtained via mapping the BTZ geometry to the
Poincar\'{e} patch of the AdS$_3$ space time as follows
\begin{equation}
  L=\frac{1}{4G_N}\log\left( \frac{\beta}{\pi \epsilon}\sinh
  \frac{2\pi x_1}{\beta}\right),
\end{equation}
where the pure time-like interval is located at $x=x_1$ distance from
the boundary and the length of both geodesics are equal since they
both lie at different time slice but at equal distance from the
boundary. Then the holographic time-like entanglement entropy for an
interval $A$ can be obtained using the RT formula as
\begin{equation}
  S_A^T=2L=\frac{c}{3}\log\left( \frac{\beta}{\pi \epsilon}
  \sinh \frac{2\pi x_1}{\beta}\right).
\end{equation}
It agrees with the corresponding dual field theory result in
\cref{fin-t-bcft-II} as $S_{bdy}=0$.

\section{Summary and discussion}\label{Summary}

To summarize, we have obtained the time-like entanglement entropy for
a pure time-like interval in the context of
AdS${}_3$/BCFT${}_2$. We first computed the time-like entanglement
entropy in a BCFT${}_2$ which involves the analytical continuation of
the standard space-like entanglement entropy to a time-like
interval. We observed that the TEE of a time-like interval at zero
temperature has three phases in contrast to the two phases of the
standard entanglement entropy in a BCFT. The new Regge phase 
is unique to the time-like interval when one end point of the
interval approaches the light cone of the mirror reflection of other
end point. We further computed the TEE of an interval on a half line
at a finite temperature and found that it has similar three
phases. The TEE has a constant and temperature independent imaginary part
in the bulk limit of small $T_0$ and in the Regge limit
of $T_0 \to 2x_1^+$, and is otherwise real.

Subsequently, we computed the time-like entanglement entropy for a
time-like interval holographically from the bulk dual to the zero and
finite temperature BCFT${}_2$. This involves the computation of RT
surface (extremal curves) in the bulk geometry. For the zero
temperature case, the bulk dual is described by a part of the AdS$_3$
geometry truncated by an EOW brane. It is observed that the holographic
TEE is described by three choices of the RT surface. The bulk and the boundary
phases consists of connected and disconnected RT surface which is
similar to the usual holographic entanglement entropy cases. The
RT surface for the Regge
phase is obtained by considering the two geodesics that go
along the plane joining  one end point
and the mirror reflection of the other end point of the interval. These two
geodesics meet at the plane perpendicular to the boundary and join together to
form the desired RT surface. We also
considered the finite temperature case where the bulk dual is BTZ
black hole cut off by a perpendicular tensionless EOW brane and observed
that it also has three phases. Interestingly, we observed that the
holographic TEE agrees precisely with the dual field theory results at both
zero and finite temperature cases in the gravity approximation.

In this paper, we find interesting phase structure for the TEE  in BCFT.
However, a basic definition of TEE directly
in terms of the field theory Hilbert space
without employing analytic continuation or holographic minimal surfaces
is lacking. It is interesting to consider concrete example of quantum
system and try to demonstrate the physical relevance of TEE.
As such, we note
that physically  the pure time interval in BCFT$_2$ describes
the configuration of a quantum dot on a half line. Our results for the TEE
should describe some kind of entanglement of a partial 
history of the dot with its complement. As the quantum dot is an easily
accessible system, one may be able to observe discontinuity for some measurables
at the transition points identified from the phase diagram of TEE.
This will give a possible demonstration of the physical relevance of the TEE.

Previously, it has been argued that the TEE in CFT
can be properly understood as a pseudo entropy. It will be interesting to consider this generalization properly to clarify the connection of TEE in a BCFT with the pseudo entropy.
It will also be extremely interesting to explore the island formalism for TEE
in the context of Island/BCFT correspondence along the lines of
\cite{Suzuki:2022xwv,Suzuki:2022yru,Izumi:2022opi}. The study of
time-like entanglement entropy is expected to shed new insights in our
understanding of the black hole interior and the emergence of
spacetime geometry from quantum entanglement. We note that the signature for an asymptotic observer
get swapped
as one passes the horizon which suggests that the time-like entropy may play an
important role in the understanding of quantum entanglement in the
interior of the black hole.
We leave these interesting issue for future investigations.

\section{Acknowledgement}
We thank Jaydeep Kumar Basak, Adrita Chakraborty and   
Dimitris Giataganas  for helpful discussion. We acknowledge
support of this work by NCTS and the grant 110-2112-M-007-015-MY3 of
the National Science and Technology Council of Taiwan.

\bibliographystyle{JHEP}
\bibliography{references}

\end{document}